\documentclass[a4paper]{PoS}
\usepackage{amsmath}
\usepackage{cite}

\newcommand{\peii}{$\pi_{e2}$}
\newcommand{\peiig}{$\pi_{e2\gamma}$}

\newcommand{\VmA}{$V$$-$$A$}  
\newcommand{\etal}{\textsl{et al.}}
\newcommand{\PR}{Phys.\ Rev.}
\newcommand{\PRL}{Phys.\ Rev.\ Lett.}
\newcommand{\RMP}{Rev.\ Mod.\ Phys.}
\newcommand{\NIM}{Nucl.\ Instrum.\ Meth.}

\title{PEN: a low energy test of lepton universality}

\ShortTitle{PEN: test of lepton universality}

\author{\speaker{Dinko Po\v{c}ani\'c},$^a$
          L.P.~Alonzi,$^a$
          V.A.~Baranov,$^b$
          W.~Bertl,$^c$
          M.~Bychkov,$^a$
          Yu.M.~Bystritsky,$^b$
          E.~Frle\v{z},$^a$
          C.J.~Glaser,$^a$
          V.A.~Kalinnikov,$^b$
          N.V.~Khomutov,$^b$
          A.S.~Korenchenko,$^b$
          S.M.~Korenchenko,$^b$
          M.~Korolija,$^d$
          T.~Kozlowski,$^e$
          N.P.~Kravchuk,$^b$
          N.A.~Kuchinsky,$^b$
          M.C.~Lehman,$^a$
          D.~Mzhavia,$^{bf}$
          A.~Palladino,$^{ac}$
          P.~Robmann,$^{g}$
          A.M.~Rozhdestvensky,$^b$
          I.~Supek,$^d$
          P.~Tru\"ol,$^g$
          A.~van~der~Schaaf,$^g$
          E.P.~Velicheva,$^b$
          M.G.~Vitz,$^a$
          V.P.~Volnykh$^b$ \quad
           (The~PEN~Collaboration)  \\
        \llap{$^a$}
          University of Virginia, Charlottesville, VA 22904-4714, USA\\ 
        \llap{$^b$}
          Joint Institute for Nuclear Research, Dubna, Moscow Region, Russia\\
        \llap{$^c$}
          Paul Scherrer Institute, Villigen-W\"urenlingen AG, Switzerland\\
        \llap{$^d$}
          Rudjer Bo\v{s}kovi\'c Institute, Zagreb, Croatia\\
        \llap{$^e$}
          Instytut Problem\'ow J\k{a}drowych im.\ Andrzeja
                 So{\l}tana, 
                 \'Swierk, Poland\\
        \llap{$^f$}
          Institute for High Energy Physics, Tbilisi State
                 University, 
                 Tbilisi, Georgia \\
        \llap{$^g$}Physik-Institut, Universit\"at Z\"urich, 
                 Z\"urich, Switzerland \\
        E-mail: \email{pocanic@virginia.edu}}

\abstract{Allowed charged $\pi$ meson decays are characterized by simple
  dynamics, few available decay channels, mainly into leptons, and
  extremely well controlled radiative and loop corrections.  In that
  sense, pion decays represent a veritable triumph of the standard model
  (SM) of elementary particles and interactions.  This relative
  theoretical simplicity makes charged pion decays a sensitive means for
  testing the underlying symmetries and the universality of weak fermion
  couplings, as well as for studying pion structure and chiral dynamics.
  Even after considerable recent improvements, experimental precision is
  lagging far behind that of the theoretical description for pion
  decays.  We review the current state of experimental study of the pion
  electronic decay $\pi^+ \to e^+\nu_e(\gamma)$, or $\pi_{e2(\gamma)}$,
  where the $(\gamma)$ indicates inclusion and explicit treatment of
  radiative decay events.  We briefly review the limits on non-SM
  processes arising from the present level of experimental precision in
  $\pi_{e2(\gamma)}$ decays.  Focusing on the PEN experiment at the Paul
  Scherrer Institute (PSI), Switzerland, we examine the prospects for
  further improvement in the near term.}

\FullConference{XIII International Conference on Heavy Quarks and Leptons\\
		22-27 May 2016\\
		Blacksburg, Virginia, USA}

\begin{document}

\section{Motivation}

$\pi$ mesons (pions), the lightest hadrons, occupy a special place for
both the weak and the strong interactions, and remain subjects of study
almost 70 years after their discovery \cite{Lat47}.  Charged pion decays
have provided an important early testing ground for the weak interaction
and radiative corrections during the development of modern particle
theory.  Decays of the charged pion proceed via the weak interaction,
strongly reflecting the properties and dynamics of the latter.  In
particular, the failure of early searches to observe the direct
electronic decay of the pion ($\pi\to e\nu$, or \peii) led to an intense
examination of the nature of the weak interaction.  A low branching
fraction of $\sim 1.3\times 10^{-4}$ was predicted \cite{Fey58} even
before the decay's discovery \cite{Faz59}.  It is a direct consequence
of the \VmA\ nature of the weak interaction, through helicity
suppression of the right-handed state of the electron.  Furthermore, the
predicted radiative corrections for the \peii\ decay \cite{Ber58,Kin59}
received quick experimental confirmation \cite{And60,DiC64}.

Pion decays have more recently been described with extraordinary
theoretical precision.  Thanks to the underlying symmetries and the
associated conservation laws, the more complicated, and thus more
uncertain, hadronic processes are suppressed.  If the experimental
results reach a precision comparable to that of their theoretical
description, pion decays offer an outstanding, clean testing ground of
universality of lepton and quark couplings.  A statistically significant
deviation from the standard model expectations would indicate the
presence of processes or interactions not included in the standard
model, affecting pion decays through loop diagrams.

\subsection{The electronic decay, $\pi^+\to e^+\nu_e$ (or \peii)
     \label{sec:pie2_motiv}} 

The $\pi^- \to \ell\bar{\nu}_\ell$ (or, $\pi^+\to \bar{\ell}\nu_\ell$)
decay connects a pseudoscalar $0^-$ state (the pion) to the $0^+$
vacuum.  At the tree level, the ratio of the $\pi \to e\bar{\nu}$ to
$\pi \to \mu\bar{\nu}$ decay widths is given by \cite{Fey58,Bry82}
\begin{equation}
    R_{e/\mu,0}^\pi \equiv \frac{\Gamma(\pi \to  e\bar{\nu})}
          {\Gamma(\pi \to  \mu\bar{\nu})}
       = \frac{m_e^2}{m_\mu^2}\cdot
        \frac{(m_\pi^2-m_e^2)^2}{(m_\pi^2-m_\mu^2)^2}
      \simeq 1.283 \times 10^{-4}\,.  \label{eq:pi_e2_tree}
\end{equation}
The first factor in the above expression, the ratio of squared lepton
masses for the two decays, comes from the helicity suppression by the
\VmA\ lepton-$W$ boson weak couplings.  If, instead, the decay could
proceed directly through the pseudoscalar current, the ratio
$R_{e/\mu}^\pi$ would reduce to the second, phase-space factor, or
approximately 5.5.  A more complete treatment of the process includes
$\delta R_{e/\mu}^\pi$, the radiative and loop corrections, and the
possibility of lepton universality violation, i.e., that $g_e$ and
$g_\mu$, the electron and muon couplings to the $W$, respectively, may
not be equal:
\begin{equation}
    R_{e/\mu}^\pi \equiv \frac{\Gamma(\pi \to  e\bar{\nu}(\gamma))}
          {\Gamma(\pi \to  \mu\bar{\nu}(\gamma))}
      = \frac{g_e^2}{g_\mu^2}\frac{m_e^2}{m_\mu^2}
        \frac{(m_\pi^2-m_e^2)^2}{(m_\pi^2-m_\mu^2)^2}
           \left(1+\delta R_{e/\mu}^\pi \right)\,,
    \label{eq:pi_e2_general}
\end{equation}
where the ``$(\gamma)$'' indicates that radiative decays are fully
included in the branching fractions.  Steady improvements over time of
the theoretical description of the $\pi_{e2}$ decay have produced
greatly refined calculations of the SM prediction, culminating at the
precision level of 8 parts in $10^5$:
\begin{equation}                           \label{eq:pi_e2_full_SM}
     \left(R_{e/\mu}^{\pi}\right)^{\rm SM} = 
       \frac{\Gamma(\pi \to e\bar{\nu}(\gamma))}
          {\Gamma(\pi \to  \mu\bar{\nu}(\gamma))}\bigg|_{\rm {calc}} =
   \begin{cases}
    1.2352(5) \times 10^{-4} & \mbox{\ \cite{Mar93},} \\
    1.2354(2) \times 10^{-4} & \mbox{\ \cite{Fin96},} \\
    1.2352(1) \times 10^{-4} & \mbox{\ \cite{Cir07}.} 
   \end{cases} 
\end{equation}
Comparison with equation (\ref{eq:pi_e2_tree}) indicates that the
radiative and loop corrections amount to almost 4\% of
$R_{e/\mu}^{\pi}$.  The current experimental precision lags behind the
above theoretical uncertainties by a factor of $\sim$\,23:
\begin{equation}
   \left(R_{e/\mu}^{\pi}\right)^{\rm exp} = 
       \frac{\Gamma(\pi \to e\bar{\nu}(\gamma))}
       {\Gamma(\pi \to  \mu\bar{\nu}(\gamma))}\bigg|_{\rm exp} =
       (1.2327 \pm 0.0023) \times 10^{-4}\,, \label{eq:pi_e2_avg}
\end{equation}
dominated by measurements from TRIUMF and PSI
\cite{Bri92a,Bri92b,Cza93,Agu15}. 

Because of the large helicity suppression of the \peii\ decay, its
branching ratio is highly susceptible to small non-\VmA\ contributions
from new physics, making this decay a particularly suitable subject of
study, as discussed in, e.g.,
Refs.~\cite{Shr81,Sha82,Loi04,Ram07,Cam05,Cam08}.  This prospect
provides the primary motivation for the ongoing PEN\cite{PENweb} and
PiENu\cite{PiENuWeb} experiments.  Of all the possible ``new physics''
contributions in the Lagrangian, \peii\ is directly sensitive to the
pseudoscalar one, while other types enter through loop diagrams.  At the
precision of $10^{-3}$, $R_{e/\mu}^\pi$ probes the pseudoscalar and
axial vector mass scales up to 1,000\,TeV and 20\,TeV,
respectively\cite{Cam05,Cam08}.  For comparison, unitarity tests of the
Cabibbo-Kobayashi-Maskawa (CKM) matrix and precise measurements of
several superallowed nuclear beta decays constrain the non-SM vector
contributions to $>20\,$TeV, and scalar ones to $>10\,$TeV \cite{PDG16}.
Although scalar interactions do not directly contribute to
$R_{e/\mu}^\pi$, they can do so through loop diagrams, resulting in a
sensitivity to new scalar interactions up to 60\,TeV \cite{Cam05,Cam08}.
The subject was recently reviewed at length in Ref.~\cite{Bry11}.  In
addition, $(R_{e/\mu}^{\pi})^{\rm exp}$ provides limits on the masses of
certain SUSY partners\cite{Ram07}, and on anomalies in the neutrino
sector\cite{Loi04}.

\subsection{Radiative electronic decay $\pi\to e\nu\gamma$ (or \peiig)  
   \label{sec:pie2g_motiv}}

\begin{figure}[tb]
  \parbox{0.47\linewidth}{
    \includegraphics[width=\linewidth]{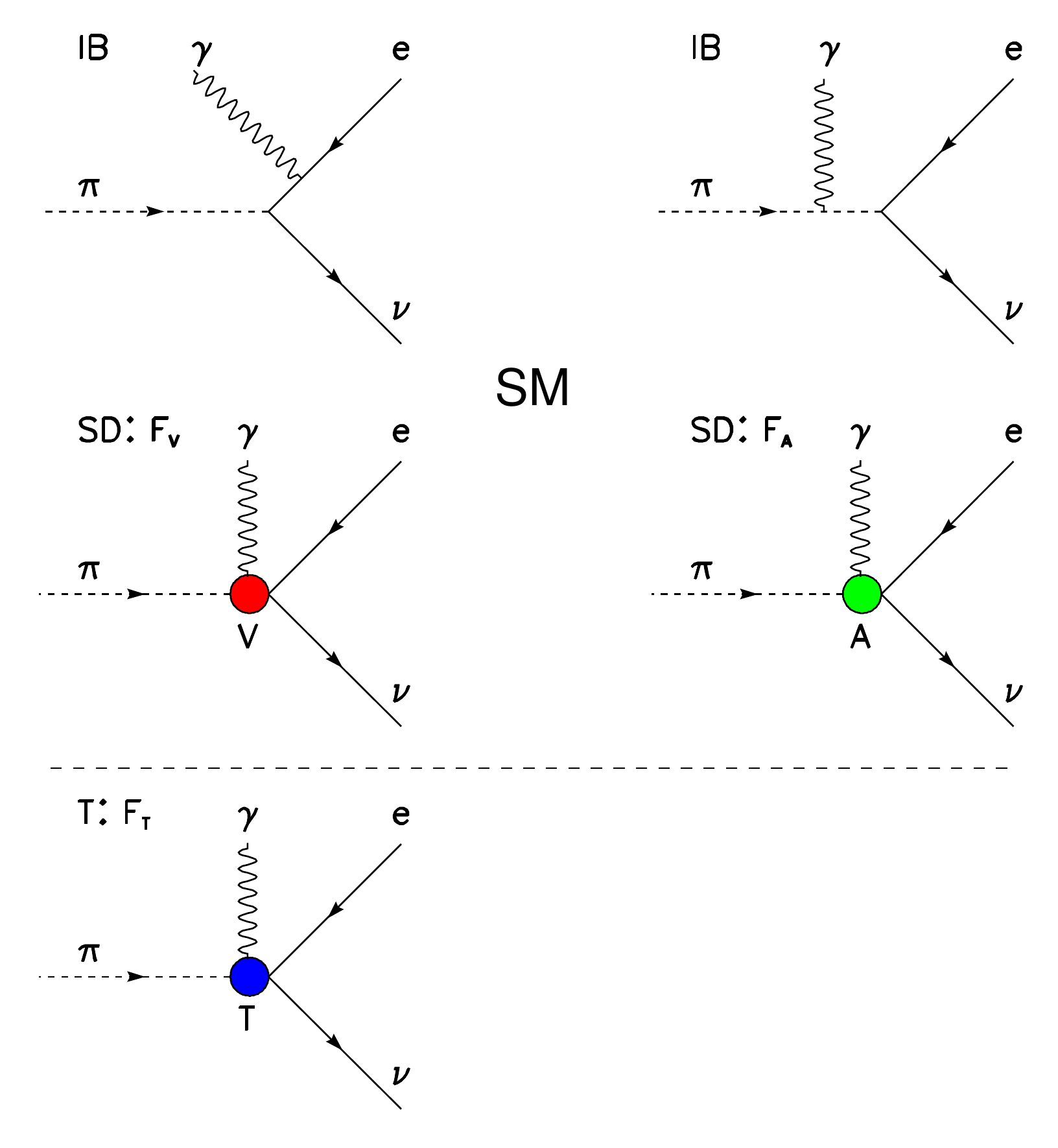}
                         }
     \hspace*{0.02\linewidth}
     \parbox{0.51\linewidth}{
        \includegraphics[width=0.49\linewidth]{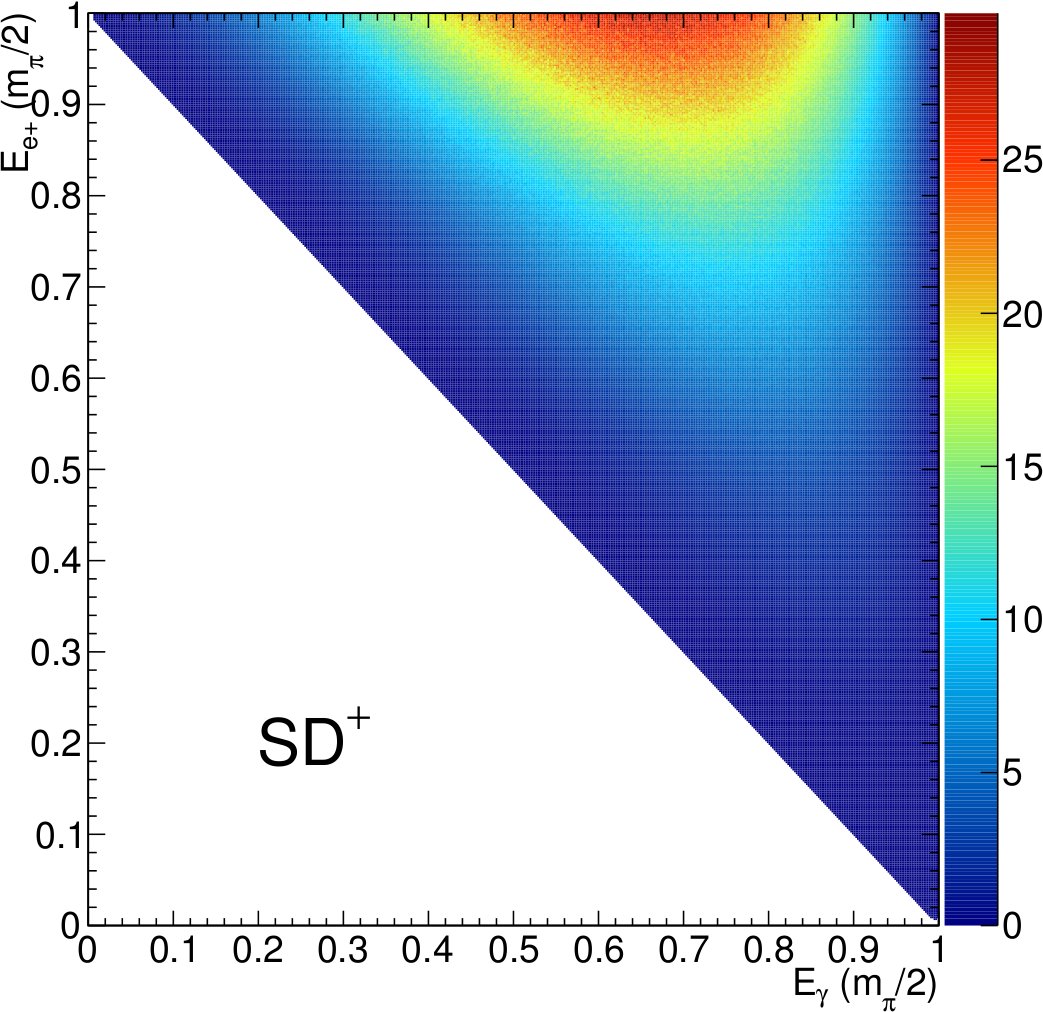}
        \hspace*{\fill}
        \includegraphics[width=0.49\linewidth]{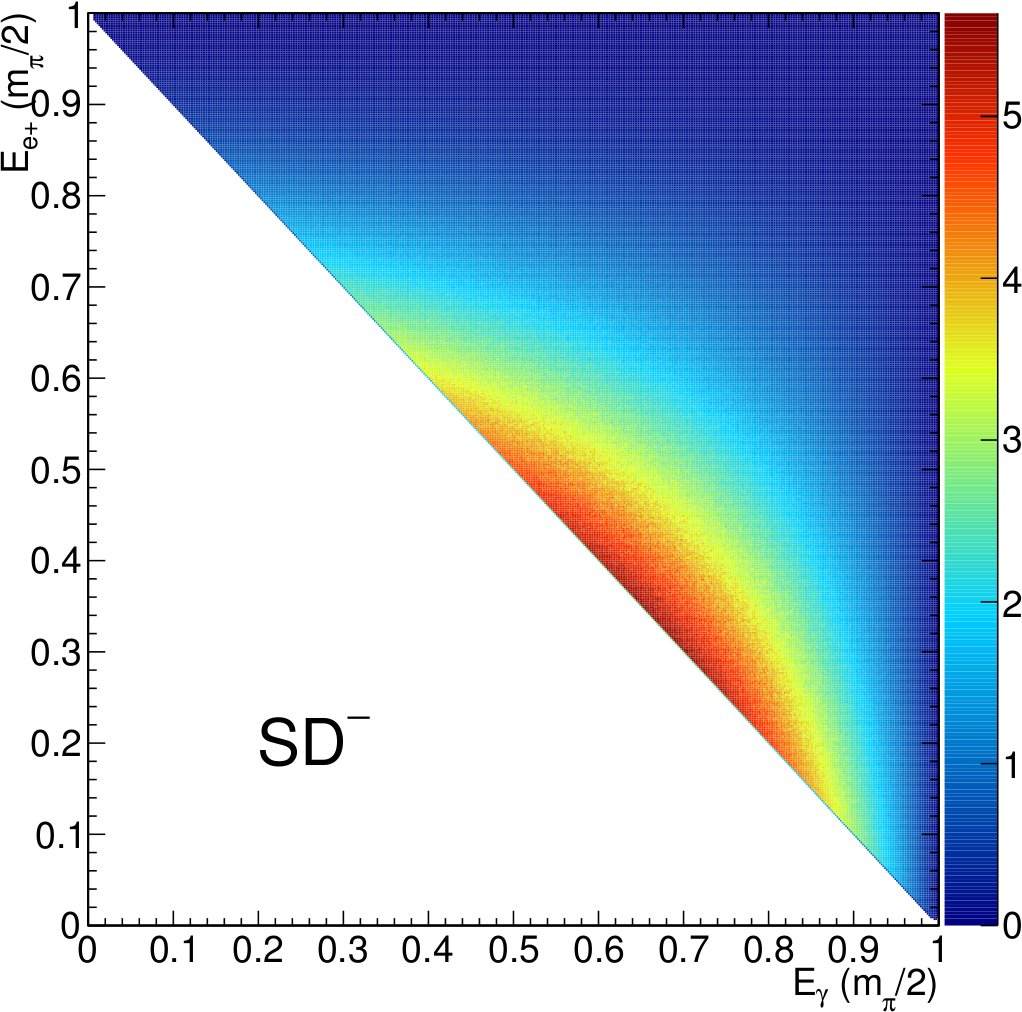} \\
        \includegraphics[width=0.49\linewidth]{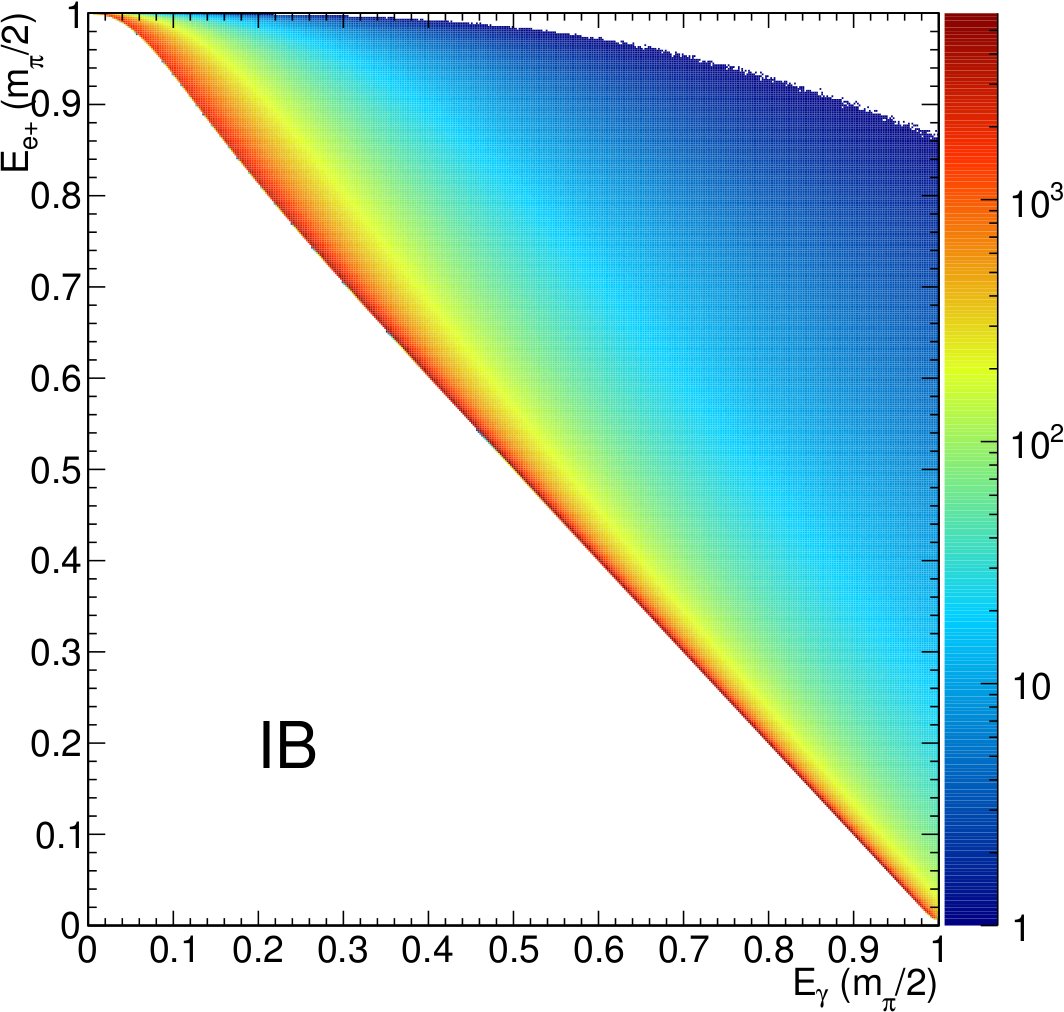}
        \hspace*{\fill}
        \includegraphics[width=0.49\linewidth]{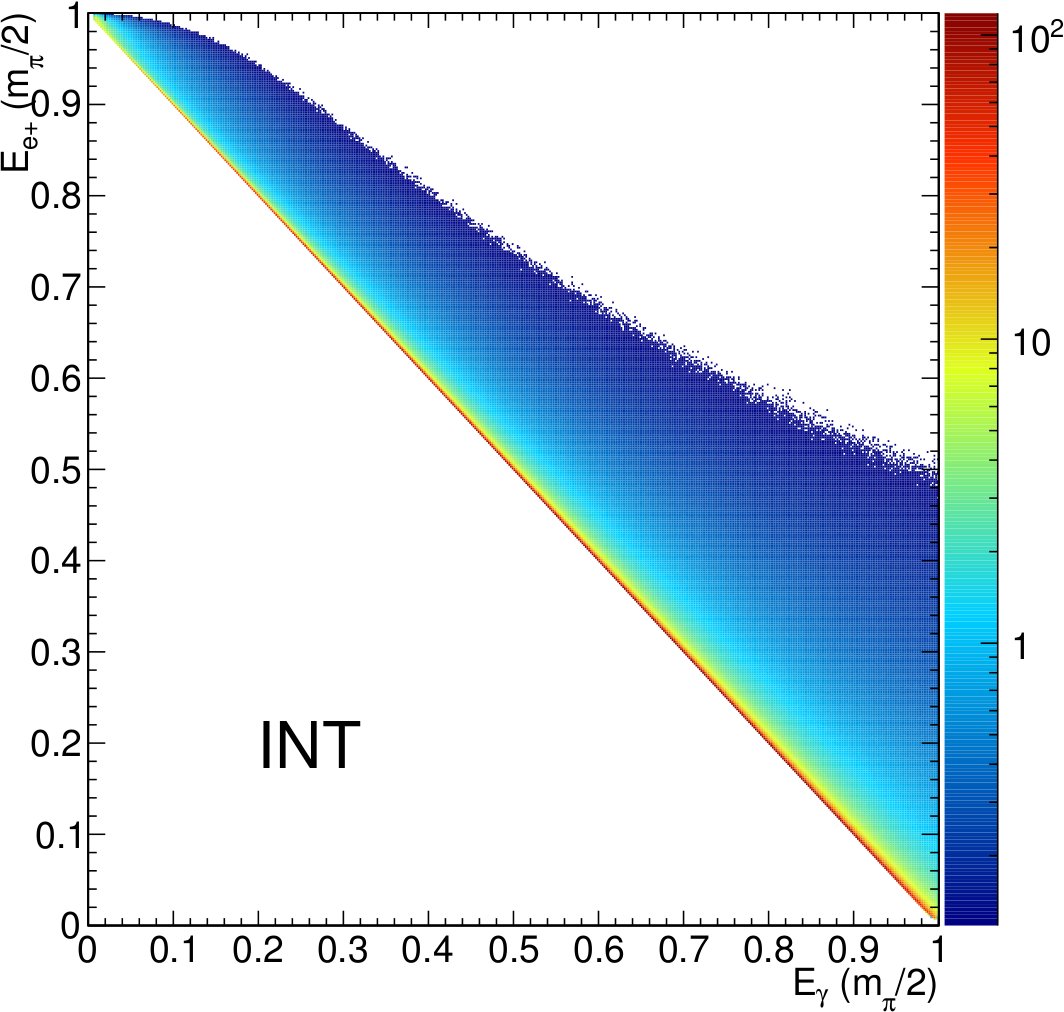}
                           }
  \caption{\emph{Left:} tree-level Feynman diagrams of the inner
    bremsstrahlung ($IB$) and structure-dependent ($SD$) amplitudes
    determined by the vector and axial-vector form factors, $F_V$ and
    $F_A$, respectively.  A new interaction type, e.g.,s one mediated by
    a hypothetical tensor particle discussed in the literature
    throughout the 1990s and 2000s \cite{Chi04}, would add an $SD$
    amplitude defined by a corresponding form factor, $F_T$.
    \emph{Right:} relative strengths of the physics amplitudes in
    $\pi\to e\nu\gamma$ decay: structure-dependent amplitudes
    $SD^+\propto (F_V+F_A)^2$ and $SD^-\propto (F_V-F_A)^2$, inner
    bremsstrahlung $IB$, and interference $INT \propto (F_V+F_A)
    S_{\text{int}}^+ + (F_V-F_A) S_{\text{int}}^-$, plotted as functions
    of $E_\gamma$ and $E_e$.}  \label{fig:pi_e2g_diags}
\end{figure}

It is impossible to discuss the \peii\ electronic decay without taking
into account its radiative variant that becomes indistinguishable in the
infrared limit $E_\gamma \to 0$.  The decay $\pi^+ \to e^+\nu_e\gamma$
proceeds via a combination of QED (inner bremsstrahlung, $IB$) and
direct, structure-dependent ($SD$) amplitudes \cite{Don92,Bry82}.  Under
normal circumstances, as in the $\pi\to\mu\nu\gamma$ decay, the direct
amplitudes are hopelessly buried under an overwhelming inner
bremsstrahlung background.  However, the strong helicity suppression of
the primary non-radiative process, $\pi\to e\nu$, also suppresses the
bremsstrahlung terms, making the direct structure dependent amplitudes
measurable in certain regions of phase space \cite{Don92,Ber13}.  (The
same helicity suppression makes sensitive searches for
non-\VmA\ interaction terms possible in precision measurements of the
primary $\pi\to e\nu$ decay, as discussed above.)  The relative
accessibility of hadronic structure amplitudes is of keen interest to
effective low-energy theories of the strong interaction, primarily
chiral perturbation theory (ChPT), which rely on the $SD$ amplitudes to
provide important input parameters.  Whereas the $IB$ amplitude is
completely described by QED, the structure-dependent amplitude can be
parametrized in terms of the pion form factors.  As seen in the
tree-level Feynman diagrams in Figure~\ref{fig:pi_e2g_diags}, standard
\VmA\ electroweak theory requires only two pion form factors, $F_A$,
axial vector, and $F_V$, vector (or polar-vector), to describe the $SD$
amplitude.
Thus, a proper experimental description of the \peiig\ decay serves
several important goals: (a) improving the accuracy of low energy
effective hadronic theories, such as ChPT, (b) enabling a precise
determination of the primary \peii\ decay rate by controlling the
systematics of the radiative decay event subset, and (c) providing the
opportunity to search for evidence of new particles with of non-SM
coupling, such as a putative tensor-interacting boson \cite{Chi04} to
which the primary, \peii\ decay process is not sensitive.  A recent
review of the subject can be found in \cite{Poc14}.

The most comprehensive study to date of the \peiig\ decay has been
performed by the PIBETA collaboration \cite{Byc09}.  This work has
provided a narrow constraint on the sum $F_V+F_A$ of the vector and
axial vector form factors of the pion, a sub-percent precision
measurement of the branching fraction for $E_\gamma > 10\,$MeV and
$\theta_{e\gamma}>40^\circ$, as well as a stringent upper bound on
$F_T$, the tensor form factor, previously the subject of considerable
controversy \cite{Poc14}.  In addition to determining $F_V$ and $F_A$
individually with greatly increased precision compared to previous
measurements, the PIBETA collaboration also evaluated, for the first
time, the $F_V$ dependence on $q_{e\nu}$, the $e^+$-$\nu_e$ invariant
mass.  In fact, the PIBETA limit on $F_T$ provides the most stringent
constraint on the strength of the tensor weak interaction \cite{Bha12},
assumed to be zero in the SM.

The PEN experiment \cite{PENweb}, discussed in more detail below,
builds on the methods, results and accomplishments of the PIBETA
collaboration.

\section{The PEN experiment at PSI}

In 2006 a new measurement of $R_{e/\mu}^{\pi}$ was proposed at the Paul
Scherrer Institute by a collaboration of seven institutions from the US
and Europe \cite{PENweb}, with the aim to reach
\begin{equation}
      {\Delta R_{e/\mu}^{\pi}}/{R_{e/\mu}^{\pi}}
       \simeq 5 \times 10^{-4} \,.    \label{eq:pen_goal}
\end{equation}
The target precision of PEN falls short of matching the theoretical
uncertainties given in equation~\ref{eq:pi_e2_full_SM}, by a factor of
about 6.  Nevertheless, PEN's target uncertainties will considerably
expand the mapped area of non-SM parameter space, as discussed in
Section \ref{sec:pie2_motiv}.  PEN has acquired data in three runs, in
2008, 2009 and 2010.

\subsection{PEN apparatus and measurement method}

\begin{figure}[tb]
    \includegraphics[width=\linewidth]{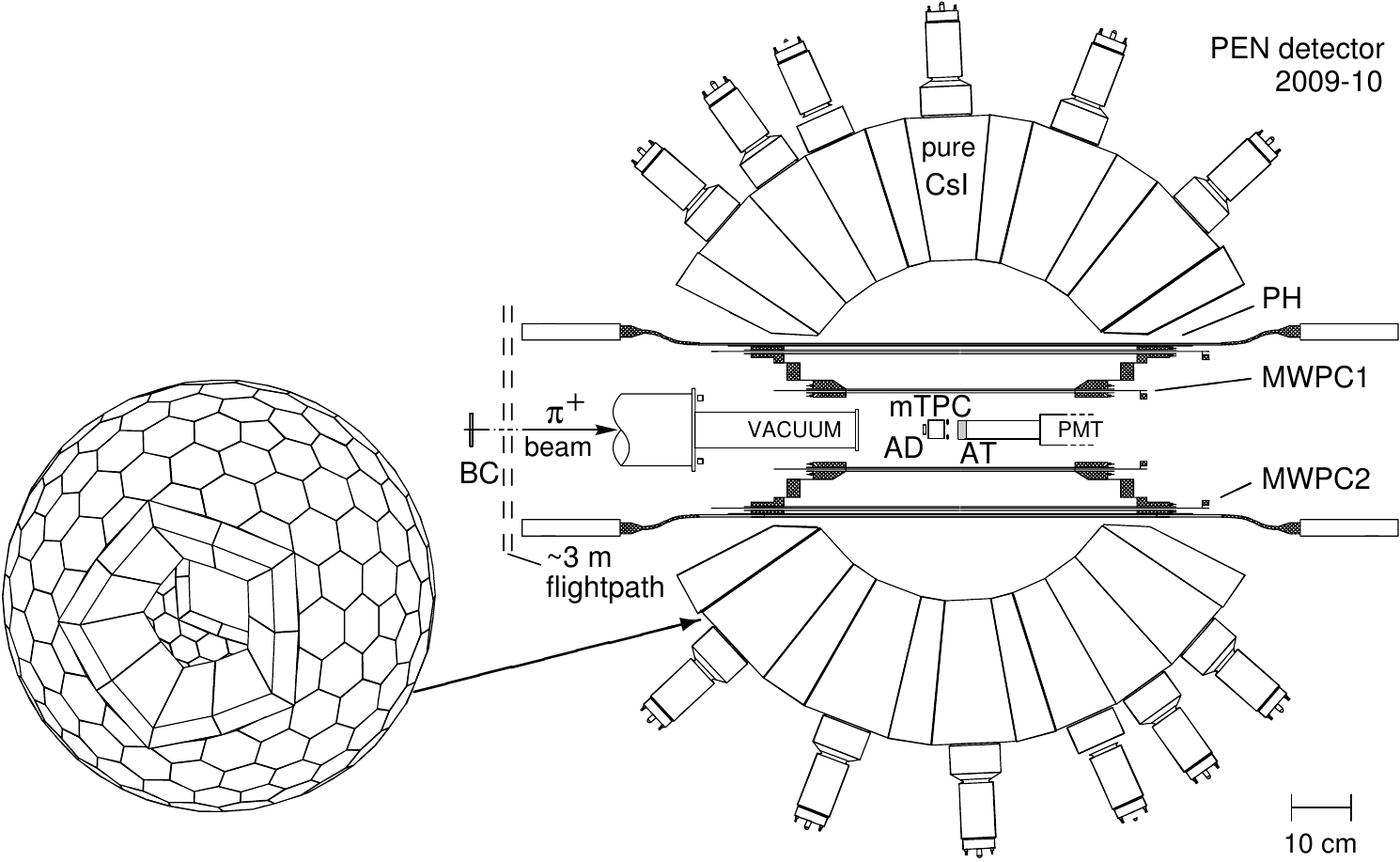}
    \caption{Schematic cross section of the PEN apparatus, shown in the
      2009-10 running configuration, with its main components: beam entry
      with the upstream beam counter (BC), 5\,mm thick active degrader
      (AD), mini time projection chamber (mTPC) followed by a passive Al
      collimator, and active target (AT), cylindrical multiwire
      proportional chambers (MWPCs), plastic hodoscope (PH) detectors
      and photomultiplier tubes (PMTs), 240-element pure CsI
      electromagnetic shower calorimeter and its PMTs.  BC, AD, AT and
      PH detectors are made of plastic scintillator.  For details
      concerning the detector performance see \cite{Frl04a}.}
      \label{fig:PEN_det}
\end{figure}

The PEN experiment uses the key components of the PIBETA apparatus with
additions and modifications suitable for a dedicated study of the
\peii\ and \peiig\ decay processes.  The PIBETA detector has been
described in detail in \cite{Frl04a}, and used in a series of
measurements of rare allowed pion and muon decay channels
\cite{Poc04,Frl04b,Byc09,Poc14}.  The major component of the PEN
apparatus, shown in Figure~\ref{fig:PEN_det},
is a spherical large-acceptance ($\sim\,3\pi$\,sr) electromagnetic
shower calorimeter.  The calorimeter consists of 240 truncated hexagonal
and pentagonal pyramids of pure CsI, 22\,cm or 12 radiation lengths
deep.  The inner and outer diameters of the sphere are 52\,cm and
96\,cm, respectively.  Beam particles entering the apparatus with
$p\simeq 75$\,MeV/$c$ are first tagged in a thin upstream beam counter
(BC) and refocused by a triplet of quadrupole magnets.  After a $\sim
3$\,m long flight path they pass through a 5\,mm thick active degrader
(AD) and a low-mass mini time projection chamber (mTPC), finally to
reach a 15\,mm thick active target (AT) where the beam pions stop.
Decay particles are tracked non-magnetically in a pair of concentric
cylindrical multiwire proportional chambers (MWPC1,2) and an array of
twenty 4\,mm thick plastic hodoscope detectors (PH), all surrounding the
active target.  The BC, AD, AT and PH detectors are all made of fast
plastic scintillator material and read out by fast photomultiplier tubes
(PMTs).  Signals from the beam detectors are sent to waveform
digitizers, running at 2\,GS/s for BC, AD, and AT, and at 250\,MS/s for
the mTPC.

\subsection{PEN data and their analysis}

\begin{figure}[b]
  \hspace*{\fill}
  \parbox{0.48\linewidth}{
          \includegraphics[width=\linewidth]{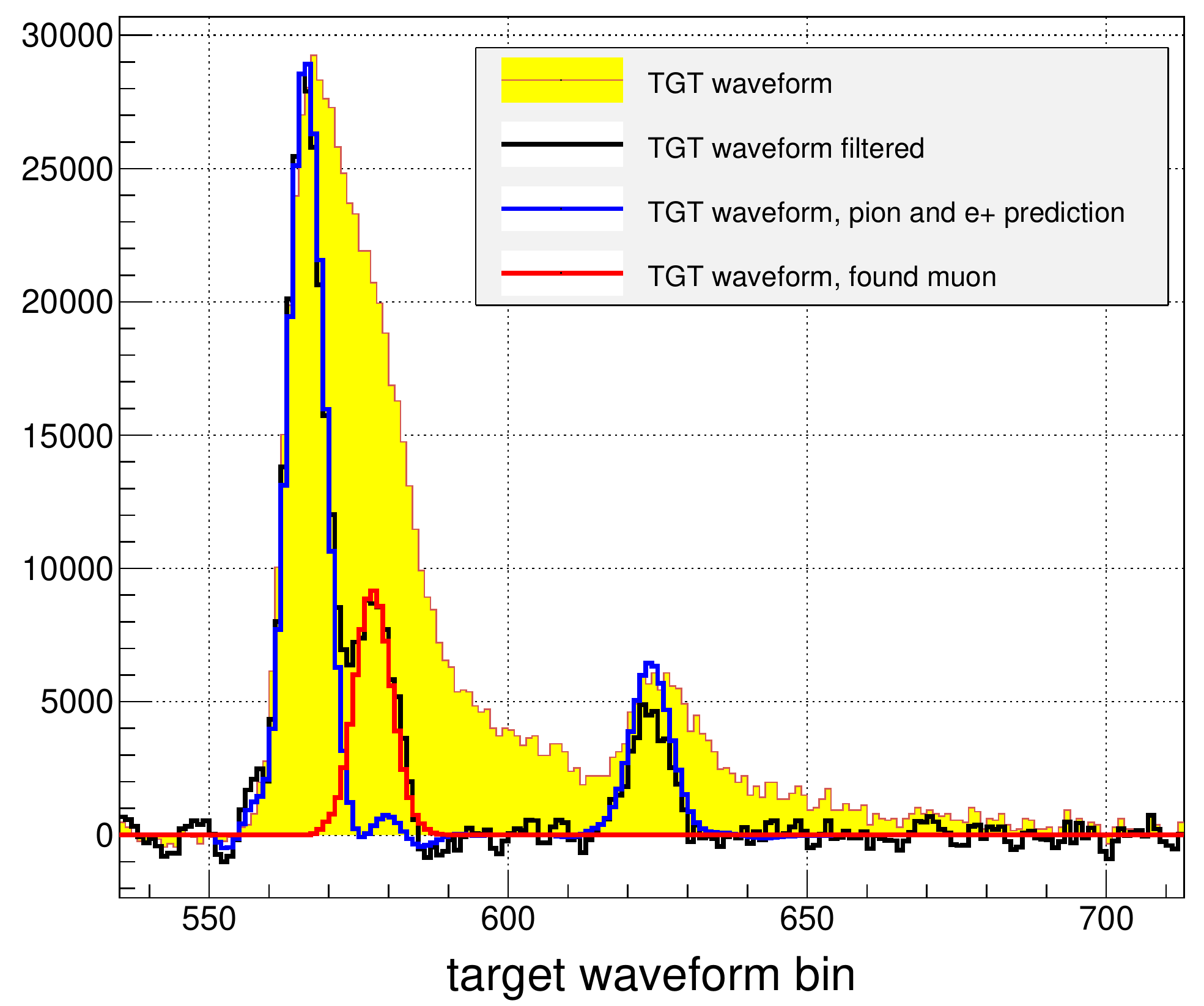}
                         }
  \parbox{0.48\linewidth}{
          \includegraphics[width=\linewidth]{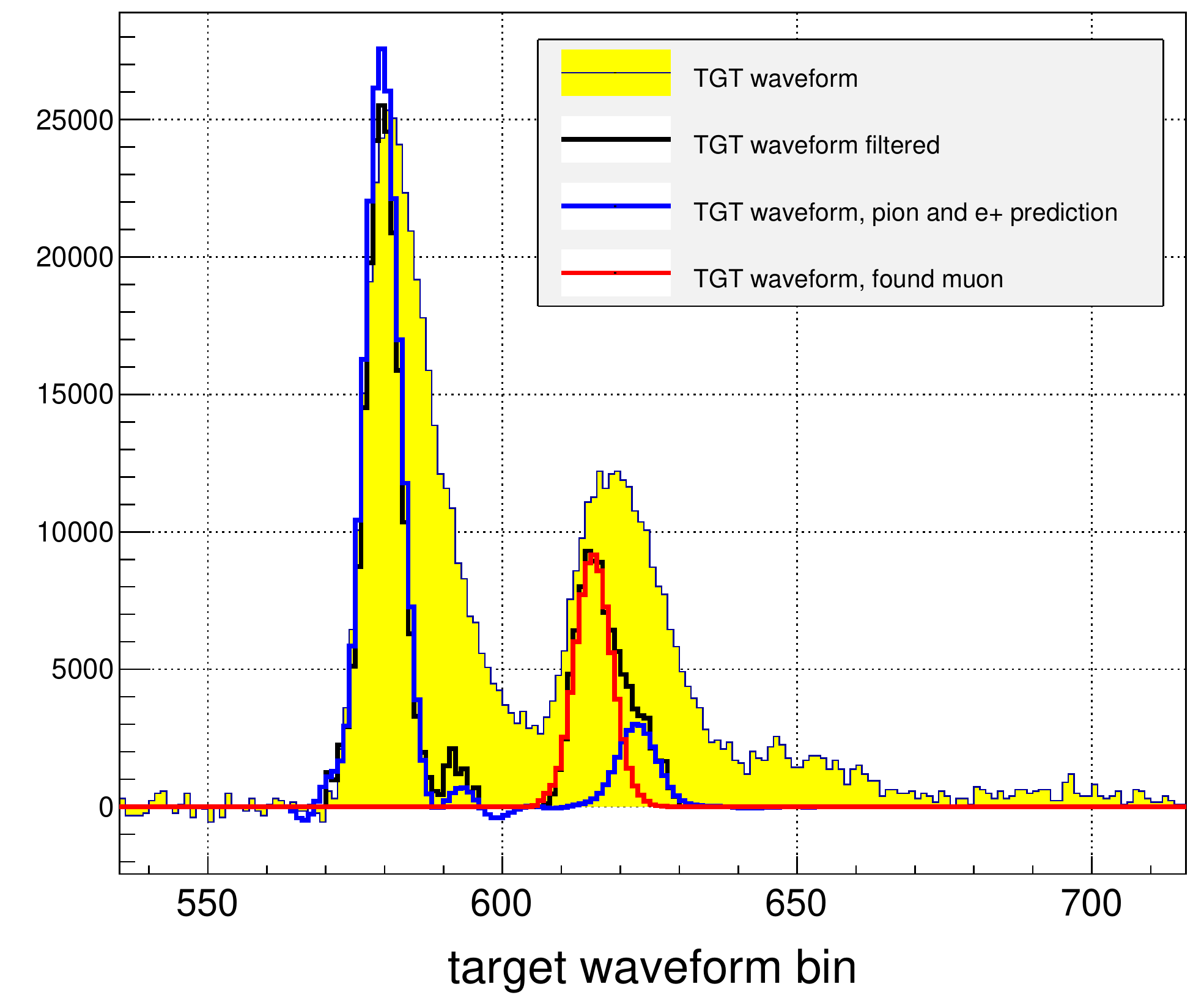}
                         } 
  \caption[waveform]{Full and filtered active target (TGT) waveform in
    the PEN experiment for two challenging $\pi\to\mu\to e$ sequential
    decay events with an early $\pi\to\mu$ decay (left) and early
    $\mu\to e$ decay (right).  The filtering procedure consists of a
    simple algebraic manipulation of the signal.  To the naked eye both
    raw waveforms appear to have two peaks only.  The separation of
    events with/without a muon signal depends critically on the accuracy
    of the predictions for the pion and positron signals.  For the pion
    the prediction is based on the times and energies observed in BC and
    AD.  For the positron the prediction depends on the PH timing and on
    the pathlength reconstructed with the pion and positron tracking
    detectors (see Figure~\ref{fig:tgt_chsq}).}
    \label{fig:wf_fits}
\end{figure}

\begin{figure}[htb]
   \hspace*{\fill}
    \includegraphics[width=\linewidth]{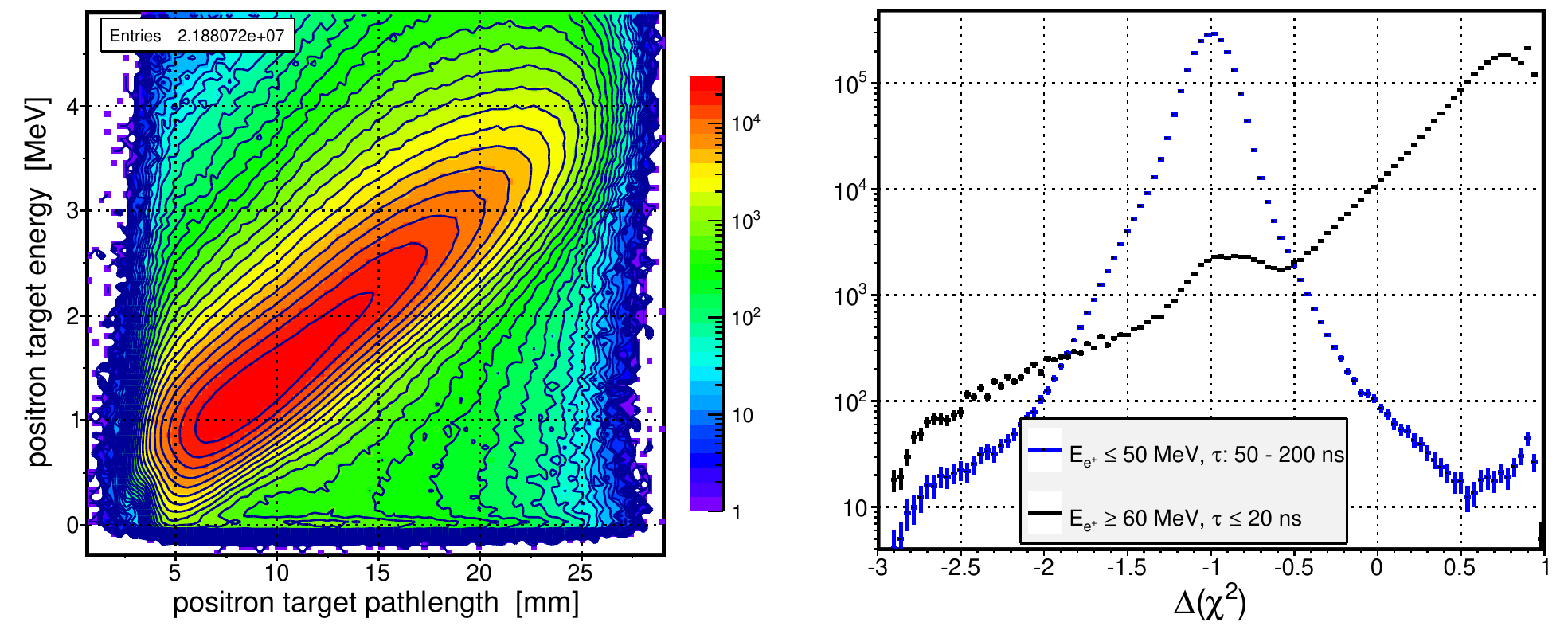}
   \caption{\emph{Left:} correlation between observed positron energy in
     the target waveform and the $e^+$ path length in the target,
     reconstructed from the observed $\pi^+$ and $e^+$ trajectories.
     Shown are events with proper $\pi\to\mu\to e$ sequences for which
     the $e^+$ signal is well separated from other signals.
     \emph{Right:} difference in $\chi^2$ for the assumptions of a
     target waveform with/without a muon pulse present.  The observable
     is normalized such that $\pi\to e\nu$ events peak at $+1$, and
     $\pi\to\mu\to e$ at $-1$.  Shown are events for two different
     combinations of $e^+$ energy and decay time resulting in almost
     pure samples of $\pi\to e\nu$ and $\pi\to\mu\to e$, respectively.
     Tiny admixtures of the other, suppressed process are readily
     identified and are of considerable help in reducing the systematic
     uncertainties. } \label{fig:tgt_chsq}
\end{figure}

Measurements of pion decay at rest, such as the one made with the PEN
detector, must deal with the challenge of separating the $\pi\to e\nu$
and $\pi\to\mu\to e$ events with great confidence.  Hence, as in earlier
\peii\ studies at rest, a key source of systematic uncertainty in PEN is
the hard to measure low energy tail of the detector response function.
The tail is caused by electromagnetic shower leakage from the
calorimeter, mostly in the form of photons.  If not properly identified
and suppressed, other physical processes contribute events to the low
energy part of the spectrum; unlike shower leakage they can also produce
high energy events.  One process is the ordinary pion decay into a muon
in flight, before the pion is stopped, with the resulting muon decaying
within the time gate accepted in the measurement.  Another is the
unavoidable physical process of radiative decay.  The latter is well
measured and properly accounted for in the PEN apparatus, as in the
PIBETA studies of \cite{Frl04b,Byc09}.  Shower leakage and pion decays
in flight can only be well characterized if the $\pi\to\mu\to e$ chain
can be well separated from the direct $\pi \to e$ decay in the target.
Therefore much effort has been devoted to digitization, filtering and
analysis of the target waveforms \cite{Pal12}, as illustrated in
Figure~\ref{fig:wf_fits}.
The method used to separate the 2-peak (\peii) and 3-peak ($\pi\to\mu\to
e$) events is illustrated and explained in Figure~\ref{fig:tgt_chsq}.
The input is provided by the beam and MWPC detectors, used to predict
the pion and positron energy depositions in the target, and the times of
their respective signals.  Once the predicted waveform is subtracted,
the remaining net waveform is scanned for the presence of a 4.1\,MeV
muon peak.  The difference between the minimum $\chi^2$ values with and
without the muon peak is reported as $\Delta(\chi^2)$, constructed so
that clean 2-peak and 3-peak fits return values of $+1$ and $-1$,
respectively.  The scan is fast and returns a $\Delta(\chi^2)$ value for
every event, as illustrated in the figure.

\begin{figure}[t]
  \hspace*{0.096\linewidth}
   \includegraphics[width=0.9\linewidth]{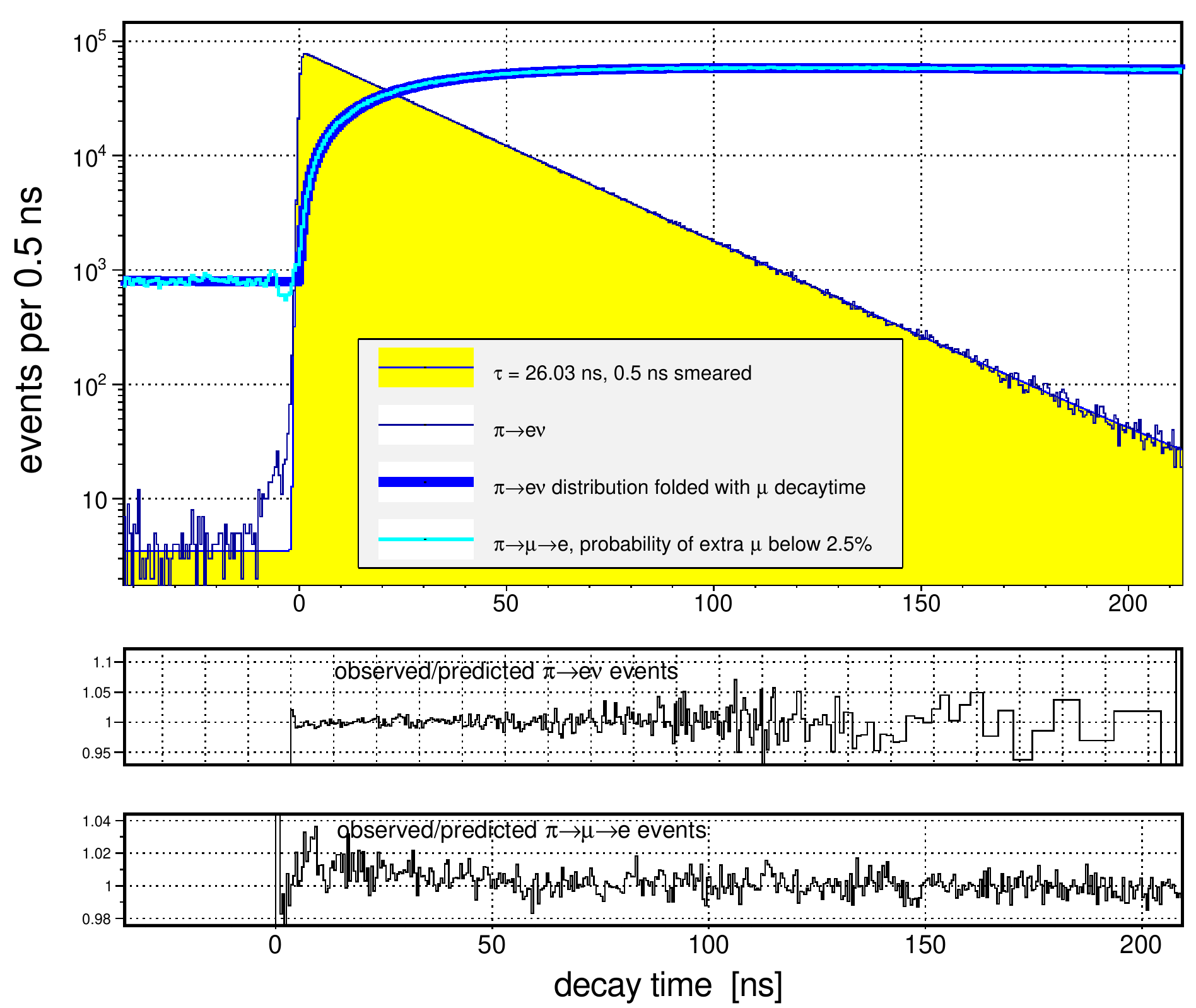}
  \caption{Decay time histograms for a subset of 2010 PEN data: $\pi\to
    e\nu$ events and $\pi\to\mu\to e$ sequential decay events.  The two
    processes are distinguished primarily by the total $e^+$ energy and
    by the absence or presence, respectively, of an extra 4.1\,MeV
    (muon) in the target due to a $\pi\to\mu$ decay.  The $\pi_{e2}$
    data are shown with a pion lifetime $\tau_\pi=26.03$\,ns exponential
    decay function superimposed.  The $\pi\to\mu\to e$ data were
    prescaled by a factor of $\sim$\,1/64; they are shown with the cut
    on the probability of $< 2.5$\% for a second, pile-up muon to be
    present in the target at $t=0$, the time of the nominal pion stop.
    The turquoise histogram gives the $\pi\to\mu\to e$ yield constructed
    entirely from the measured $\pi\to e\nu$ data folded with the $\mu$
    decay rate, and corrected for random muons; it perfectly matches the
    bold dark blue histogram.  The two lower plots show the observed to
    predicted ratios for \peii\ and $\pi\to\mu\to e$ events,
    respectively.  The scatter in the ratio plots is statistical in
    nature. } \label{fig:pen-decaytime}
\end{figure}

A particularly telling figure regarding the PEN data quality is the
decay time comparison of the $\pi\to e\nu$ decay and $\pi\to \mu\to e$
sequence, shown in Figure~\ref{fig:pen-decaytime} for a subset of data
recorded in 2010.
The $\pi\to e\nu$ data follow the exponential decay law over more than
three orders of magnitude, and perfectly predict the measured
$\pi\to\mu\to e$ sequential decay data once the latter are corrected for
random (pile-up) events.  Both event ensembles were obtained with
minimal requirements (cuts) on detector observables, none of which
biases the selection in ways that would affect the branching ratio.  The
probability of random $\mu\to e$ events originating in the target can be
controlled in the data sample by making use of multihit time to digital
converter (TDC) data that record past pion stop signals.  With this
information one can strongly suppress events in which an ``old'' muon
was present (``piled up'') in the target by the time of the pion stop
that triggered the readout.

The ``intrinsic'' low energy tail of the PEN response function below
$\sim$ 50\,MeV, due to shower losses for \peii\ decay events for pions
at rest, amounts to approximately 2\% of the full yield.  Events with
$\pi\to\mu$ decays in flight, with subsequent ordinary Michel decay of
the stopped muon in the target, add a comparable contribution to the
tail.  The two contributions can be simulated accurately, with the
respective detector responses independently verified through comparisons
with measured data in appropriately selected processes and regions of
phase space.
Although verified through comparisons with Monte Carlo simulations, the
intrinsic tail itself is not directly measurable at the required
precision because of the statistical uncertainties arising in the tail
data selection procedure.  Radiative decay processes are directly
measurable and accounted for in the branching fraction data analysis.
More information about the PEN/PIBETA detector response functions is
given in \cite{Frl04a}.

We next turn our attention to the strongly radiative decay events, i.e.,
those with energetic photons that lead to clearly separated $e^+$ and
$\gamma$ initiated showers in the PEN CsI calorimeter.  The PEN
collaboration has recorded a substantial new data set of such events, to
be added to the previously recorded PIBETA \peiig\ data set.  Since the
PEN beam stopping rate is lower than that used in the cleanest, 2004
PIBETA run, the impact on the statistical uncertainties, while
significant, will not be profound.  However, the cleaner running
conditions of PEN allow for easier access to the kinematic region
$E_\gamma,E_e < 50$\,MeV, previously strongly contaminated by the muon
decay background.  These are the regions needed to probe the $SD^-$
amplitude critically, as seen in Figure~\ref{fig:pi_e2g_diags}.
Accessing $SD^-$ opens the prospects for a new determination of the
quantity $F_V-F_A$, poorly constrained in the main PIBETA result
\cite{Byc09}, and, hence, for an improved model-independent experimental
determination of $F_V$.  These results will be forthcoming in the near
future.

\section{Conclusions}

During the three production runs, from 2008 to 2010, the PEN experiment
accumulated some $2.3 \times 10^7$ $\pi\to e\nu$, and more than $1.5
\times 10^8$ $\pi\to\mu\to e$ events, as well as significant numbers of
pion and muon radiative decays.  A comprehensive blinded analysis is
under way to extract a new experimental value of $R_{e/\mu}^{\pi}$.  As
of this writing, there appear to be no obstacles that would prevent the
PEN collaboration from reaching a precision of $\Delta R/R < 10^{-3}$.
The competing PiENu experiment at TRIUMF \cite{PiENuWeb} has a similar
precision goal.  The near to medium future will thus bring about a
substantial improvement in the limits on $e$-$\mu$ lepton universality,
and in the related limits on non-SM, non-\VmA\ processes and couplings.

It is important to note that even subsequent to the completion of the
PEN and PiENu data analyses, there will remain considerable room for
improvement of experimental precision with high payoff in terms of
limits on physics not included in the present Standard Model.  If fully
successful, the current experiments will bridge the current gap between
the experimental and theoretical uncertainty levels only half-way,
leaving room for new experiments.  Furthermore, this work remains
relevant and complementary to the direct searches on the energy frontier
currently underway at particle colliders, providing valuable theoretical
cross checks.

\acknowledgments

This work has been supported by grants PHY-0970013 and PHY-1307328 from
the United States National Science Foundation.

\end{document}